%% file: main.tex
\documentclass{article}
\usepackage{spconf,amsmath,graphicx}
\usepackage{multirow}
\usepackage{subcaption}
\usepackage[table,xcdraw]{xcolor}
\usepackage{enumitem}
\usepackage{url}
\usepackage{hyperref}

\title{FedKPer: Tackling Generalization and Personalization in Medical Federated Learning via Knowledge Personalization}
\name{Anonymous Authors}
\address{Anonymous Affiliations}
\name{Zoe Fowler and Ghassan AlRegib}
\address{Georgia Institute of Technology}

\begin{document}
\clearpage
\onecolumn
\begingroup
\sloppy
\emergencystretch=3em
\input{cover} 
\endgroup
\clearpage
\twocolumn
%
\maketitle
\input{Sections/0_abstract}
\input{Sections/1_intro}
\input{Sections/2_litreview}
\input{Sections/3_method}
\input{Sections/4_results}
\input{Sections/5_conclusion}
\section{Acknowledgments}
\label{sec:acknowledgments}
This material is based upon work supported by the National Science Foundation Graduate Research Fellowship under Grant No. DGE-2039655 and award number 2515189.
\vfill\pagebreak

\bibliographystyle{IEEEbib}
\bibliography{strings,refs}

\end{document}

%% file: cover.tex





\onecolumn

\begin{description}[
  leftmargin=2.8cm,
  labelwidth=2.3cm,
  labelsep=0.4cm,
  align=parleft,
  style=multiline,
  font=\normalfont
]

\item[\textbf{Citation}]{Z. Fowler and G. AlRegib, "FedKPer: Tackling Generalization and Personalization in Medical Federated Learning via Knowledge Personalization," in IEEE International Conference on Image Processing (ICIP), April 2026.}


\item[\textbf{Review}]{Date of acceptance: 30 April 2026}

\item[\textbf{Data and Codes}]{\url{https://github.com/olivesgatech/FedKPer}}

\item[\textbf{Bib}]{@inproceedings\{fowler2026fedkper,\\
author=\{Z. Fowler and G. AlRegib\},\\
booktitle=\{IEEE International Conference on Image Processing\},\\
title=\{FedKPer: Tackling Generalization and Personalization in Medical Federated Learning via Knowledge Personalization\},\\
year=\{2026\}\\\}
}


\item[\textbf{Contact}]{
\href{mailto:alregib@gatech.edu}{alregib@gatech.edu}\\
\url{https://alregib.ece.gatech.edu/}%
}

\end{description}

\thispagestyle{empty}
\clearpage
\setcounter{page}{1}


%% file: Sections/0_abstract.tex
\begin{abstract}
Federated learning (FL) holds great potential for medical applications. However, statistical heterogeneity across healthcare institutions poses a major challenge for FL, as the global model struggles both to generalize across unseen patient populations and to adapt to the unique data distributions of individual hospitals. This heterogeneity also exacerbates forgetting at both the global and local level, resulting in previous learned patient patterns to be misclassified after model updates. While prior work has largely treated generalization and personalization as separate challenges, we show that a better balance between the two can be achieved through selective alignment with the global model and a modified aggregation scheme, which together mitigate the effects of statistical heterogeneity. Specifically, we introduce FedKPer, which introduces knowledge personalization into the training stage of each local device. Afterwards, generalization is considered via the global model aggregation process, where local updates that are reliable and label-diverse are emphasized. We evaluate the performance of FedKPer, devising additional metrics that relate to common consequences of forgetting. Overall, we demonstrate FedKPer improves the generalization-personalization trade-off without sacrificing retention.
\end{abstract}
\begin{keywords}
federated learning, personalization, generalization, medical machine learning
\end{keywords}

%% file: Sections/1_intro.tex
\section{Introduction}
In recent years, advances in technology and evolving patient needs have transformed the medical field, with machine learning showing promise across numerous applications. Yet, translating these approaches into clinical practice remains difficult. Robust medical models require diverse data from multiple institutions to avoid overfitting and institutional bias, but assembling large centralized datasets is often infeasible due to privacy regulations \cite{sheller2020federated}. 
Motivated by the challenge to address both generalization and patient privacy concerns, federated learning (FL) has emerged as a promising solution by enabling collaborative training across institutions without sharing raw patient data \cite{mcmahan2017communication, rieke2020future}.

FL is a distributed training paradigm, where local devices, like hospitals, share only model parameters with a central server. The server then aggregates these parameters to form a global model \cite{mcmahan2017communication}. FL algorithms are typically evaluated in two ways: generalization, measured by the global model’s performance on a held-out test set, and personalization, measured by how well local models adapt to their own data \cite{mcmahan2017communication, firdaus2023personalized}.

A core challenge in FL is statistical heterogeneity, i.e. non-IID data across local devices, which is a natural consequence in medical contexts where patient populations vary widely across sites.
This heterogeneous data results in a drastic decrease in the \textit{both} the generalization and personalization performance of standard FL algorithms \cite{firdaus2023personalized, zhao2018federated}. 
Despite the challenge of statistical heterogeneity existing at both the global and local level, most prior work focuses on either the generalization or the personalization challenge instead of considering both.
Classical FL approaches like FedAvg prioritize global model generalization, but this often comes at the cost of local accuracy \cite{mcmahan2017communication}. Personalized federated learning (pFL), on the other hand, improves local performance, ultimately sacrificing generalization capabilities \cite{tan2022towards}.

\begin{figure}
    \centering
    \includegraphics[width=0.45\textwidth]{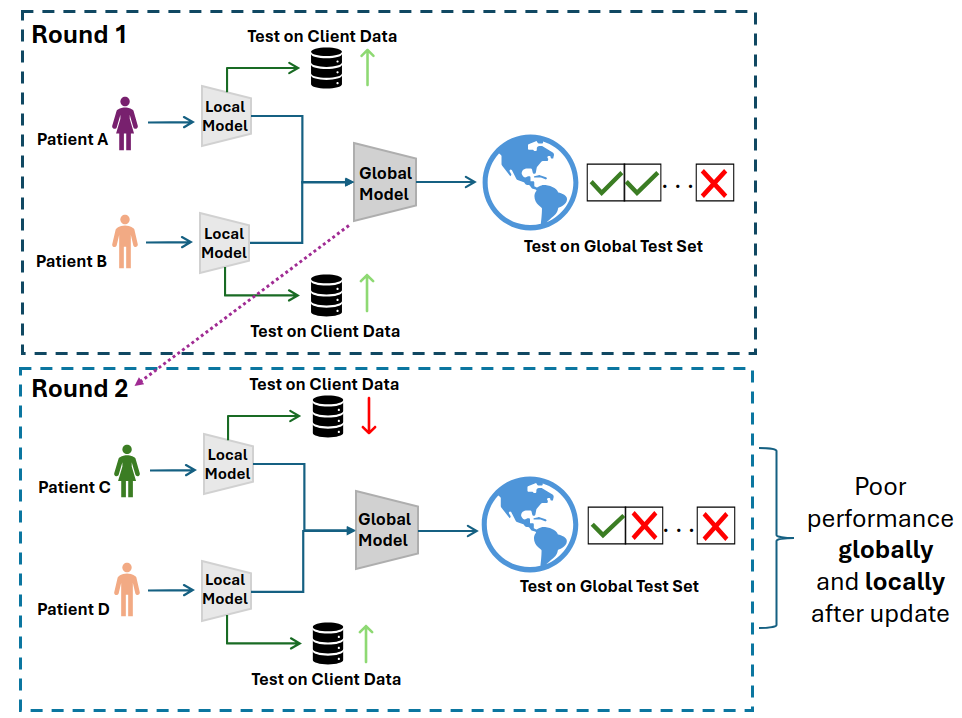}
    \caption{Statistical heterogeneity, combined with partial client participation, influences both \textit{personalization} and \textit{generalization}.}
    \label{fig:intro}
\end{figure}

Additionally, most FL work evaluates methods primarily by accuracy, overlooking other important behaviors. One is forgetting: under statistical heterogeneity (and partial client participation), updates can overwrite previously acquired knowledge, so performance on earlier clients/classes degrades over rounds and ultimately reduces the global model’s ability to generalize consistently across the population \cite{lee2022preservation}. Knowledge distillation is often used to curb this drift by transferring global predictions to local models. However, with heterogeneity and partial sampling, the global model may be a weak teacher for many clients; enforcing its logits can misguide local training and produce poorly adapted local models. Aggregating these updates can then deteriorate the global model (Figure~\ref{fig:intro}). Thus, blind distillation can simultaneously constrain personalization and harm global generalization.

The previous dilemma highlights the need for establishing a trade-off between retaining prior knowledge and learning new client information. 
Hence, in this paper, we present FedKPer, a solution that addresses the generalization and personalization challenge by adaptively balancing this trade-off. 
We propose \textit{knowledge personalization}, where we personalize the knowledge learned by each local client during its training stage. After training, we modify the model aggregation strategy to be influenced by local models that are both reliable and label-diverse, increasing the generalization. 
We also introduce additional evaluation metrics for FL algorithms, rooted in forgetting. Rather than optimizing personalization alone, FedKPer targets the clinically-relevant regime where models must both adapt locally and remain transferable across institutions. Accordingly, our goal is not to maximize per-client accuracy in isolation, but to improve the trade-off between personalization and generalization, while minimizing forgetting. Experimental results demonstrate the effectiveness of FedKPer at improving this trade-off, establishing a better balance between the generalization and personalization tasks on medical image datasets.

%% file: Sections/2_litreview.tex
\section{Related Works}
Many prior approaches have aimed to tackle the heterogeneity problem prevalent in FL from the standpoint of generalization. FedProx \cite{li2020federated}, for instance, adds a proximal term to the local training stage of each local client. Other studies have attempted to understand heterogeneity from a forgetting lens, such as FedCurv, which compels local models towards finding a shared optimum \cite{shoham2019overcoming}. Similarly, FedNTD \cite{lee2022preservation} aims to preserve the global model's perspective on not-true classes on the local data to alleviate forgetting in a framework inspired by knowledge distillation. pFL, on the other hand, tackles the heterogeneity challenge by developing tailored local models. Methods like FedALA \cite{zhang2023fedala} selectively download information from the global model, while methods like FedKD \cite{wu2022communication} uses the concept of knowledge distillation to learn a small homogeneous student model.

Prior work has explored balancing generalization and personalization in FL. FedBABU, for instance, learns a shared representation and then fine-tunes a personalized head \cite{oh2021fedbabu}. While effective, this design separates the two objectives into distinct stages. FedKPer, instead, couples them end-to-end by controlling what knowledge is trusted (client-side) and which updates dominate (server-side), optimizing the generalization-personalization trade-off rather than either metric alone.



%% file: Sections/3_method.tex
\section{Methodology}
\begin{figure*}
    \centering
    \includegraphics[width=0.68\textwidth]{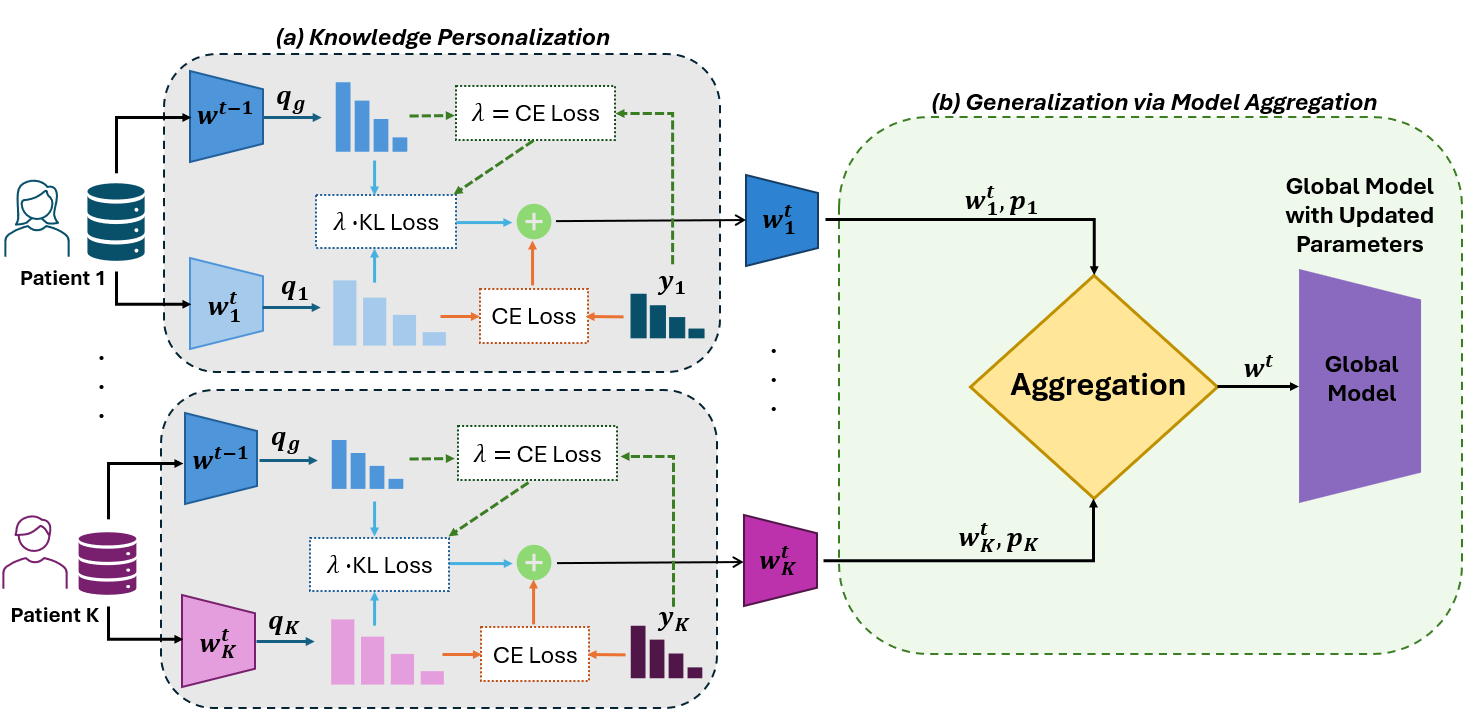}
    \caption{FedKPer introduces \textbf{(a)} Knowledge Personalization via adaptive regularization with the global model's understanding of the local data and \textbf{(b)} enhanced generalization via modifying the aggregation stage from local site and model statistics}
    \label{fig:method}
\end{figure*}
\subsection{Federated Learning Setup}
In standard FL setups, each of the $K$ total clients hold their own local dataset $\mathcal{D}_{k}$ of size $n_{k}$ \cite{mcmahan2017communication}. 
At communication round $t$, the server samples a subset of clients $K^{t}$ and broadcasts the current global model $w^{t-1}$. Federated learning overall seeks to minimize the global objective
\begin{equation}
\begin{aligned}
    \min_{w}\; F(w) \;=\; \sum_{k=1}^{K} p_k F_k(w), \\
    \text{with }
    F_k(w) = \frac{1}{n_{k}}\sum^{n_{k}}_{i=1}\ell(x_{i},y_{i};w),
    \end{aligned}
\end{equation}
where $p_k = \frac{n_k}{\sum_{j=1}^K n_j}$ and $\ell(\cdot)$ denotes the loss on client $k$'s local data.
Each selected client $k\in K^t$ approximately minimizes $F_k(w)$ starting from $w^{t-1}$ to obtain updated parameters $w_k^{t}$.
The server then aggregates the received client models to form the new global model:
\begin{equation}
\label{eq:fedavg}
w^{t} \;=\; \sum_{k\in K^t} \frac{n_k}{\sum_{j\in K^t} n_j}\, w_k^{t}.
\end{equation}

\subsection{Formulation of FedKPer}
Our proposed method FedKPer (Figure \ref{fig:method}) is made up of two distinct components: (1) knowledge personalization at the local training stage and (2) regularization of the global model from this personalization via a novel aggregation strategy. We describe each of these elements below.

\textit{Knowledge Personalization } 
Vanilla knowledge distillation (KD) encourages a client model to match the global model’s logits on local data, but the global model may be unreliable under heterogeneity. Thus, we modify the local training stage to introduce the new loss function $\ell$ as follows: 
\begin{equation}
\label{eqn_loss}
    \ell = \ell_{\text{CE}}(q_{k},y) + \lambda \ell_{\text{KD}}(q_g, q_k)
\end{equation}
where $\ell_{\text{CE}}(q_{k},y)$ is cross-entropy between the client predictions $q_k$ and labels $y$, and $\ell_{\text{KD}}$ is KL divergence-based distillation on the client data.
Within this loss, we introduce an adaptive reliability weight $\lambda$ on the distillation term. As indicated by Figure \ref{fig:method}.a, we set $\lambda = \frac{1}{\ell_{\text{CE}}(q_{g},y)}$ such that the influence of the global model’s predictive distribution is suppressed when it is inaccurate on a client’s data. In other words, the global model is treated as a teacher whose guidance is trusted only to the extent that it is empirically correct on the current client. This prevents harmful imitation of a globally-averaged hypothesis that may be distorted by heterogeneity. 
Consequently, this adaptive distillation yields principled personalization, letting each client retain globally transferable structure while flexibly adapting to its own label distribution. Since $\lambda$ can become large if the global model attains near-perfect accuracy (rare in label-heterogeneous scenarios), we additionally cap $\lambda$ at 10 as a numerical safeguard. Empirically, we verified that this cap has negligible effect on performance across our experiments, indicating it serves purely as a stability safeguard rather than a tuned performance parameter. Additionally, we apply gradient-norm clipping (max norm of 5) to limit the impact of occasional outlier updates.

\textit{Generalization via Aggregation } 
While FedKPer’s knowledge personalization targets each client’s ability to adapt to its local distribution via selective alignment with the global model, this client-side objective does not directly optimize the global objective. Under heterogeneous data, it may increase disagreement between client optima, so averaging updates, especially with dataset-size weights, may yield a global model that generalizes poorly.
Instead, we propose a more informative model aggregation technique that prioritizes client updates that are both reliable and label-diverse as demonstrated in Figure \ref{fig:method}.b. For each sampled client $k \in K^{t}$ at round $t$, we evaluate each local model on its corresponding local train set, obtaining accuracy $\mathcal{A}_k^{t}$ as a reliability proxy.
Additionally, we define a diversity score via the client label histogram entropy. Hence, for round $t$, client $k$'s label histogram over the $C$ total classes has entries $m_{k,c}$ and total count $M_k \;=\; \sum_{c=1}^{C} m_{k,c}$.
We then define the corresponding empirical label distribution as
\begin{equation}
\label{eqn_labeldist}
    \pi_{k,c} = \frac{m_{k,c}}{M_k + \varepsilon}
\end{equation}
Then, the normalized-entropy diversity score is
\begin{equation}
\label{eqn_div}
    d_k = \frac{-\sum_{c=1}^{C} \pi_{k,c} \log\!\big(\pi_{k,c}+\varepsilon\big)}{\log(C)+\varepsilon}
\end{equation}
We then combine performance and diversity into an unnormalized aggregation score
\begin{equation}
\label{eqn_score}
    \bar{p}_k = \big(\mathcal{A}_k^{t}\big)\big(\varepsilon + d_k\big)
\end{equation}
 where we utilize $\varepsilon = 10^{-12}$ throughout for stability. We can then write the final model aggregation weights for sampled clients $k \in K^{t}$ as $p_{k} = \frac{\bar{p}_{k}}{\sum_{j\in K^{t}}\bar{p}_{j}}$, where $\sum_{k\in K^{t}} p_{k} = 1$. In effect, this weighting limits the extent to which updates from highly skewed client label distributions can dominate the global step, encouraging the aggregated model to reflect a broader range of label groups across the participating institutions. This is important due to class imbalance, where uniformly averaging can over-emphasize clients that see only a narrow subset of classes, leading to class-biased global drift. Importantly, this aggregation requires no raw data sharing; only model parameters and the combined aggregation score ($\bar{p}_k$) for a given client are transmitted to the server.


\subsection{Forgetting-based Metrics}
\label{subsec:metrics}
\begin{figure}[h!]
    \centering
    \includegraphics[width=0.4\textwidth]{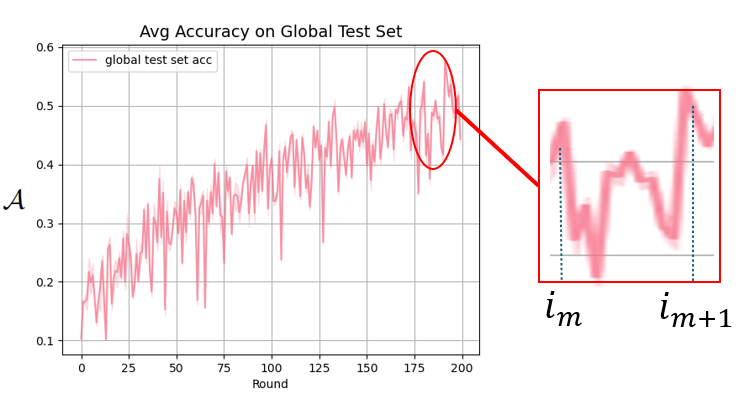}
    \caption{FedAvg: Example window of region where accuracy decreases after initial peak at arbitrary index $i_{m}$}
    \label{fig:acc_peaks}
\end{figure}

\textit{Consistency } While rarely studied or measured, forgetting can explain many undesirable properties of FL algorithms, such as inconsistent behavior in accuracy, demonstrated via Figure~\ref{fig:acc_peaks} marked by sudden drops in accuracy. 
To quantify how reliably performance is maintained after reaching a strong operating point, we partition the accuracy trajectory into consecutive \emph{recovery intervals}, each spanning from a record-high peak to the first subsequent round where accuracy recovers to (or exceeds) that peak.
Let $\{\mathcal{A}_t\}_{t=0}^{T}$ denote the accuracy over communication rounds.
We define $i_m$ as the index of the $m$-th record-high (running-maximum) peak, i.e., $\mathcal{A}_{i_m} \ge \max_{0\le i \le i_m}\mathcal{A}_i$.
Starting from $i_m$, we define the end of the corresponding recovery interval as the first index $i_{m+1} > i_m$ for which performance returns to (or exceeds) the previous peak level, $\mathcal{A}_{i_{m+1}} \ge \mathcal{A}_{i_m}$.
We scan rounds sequentially, declare a new peak whenever $\mathcal{A}_t$ exceeds the running maximum, and close the current interval at the first subsequent round where $\mathcal{A}_t$ returns to at least that peak.
For each recovery interval $[i_m, i_{m+1}]$, we define the \emph{inter-peak forgetting rate} (IPFR) as the average relative deviation from the peak:
\begin{equation}
\label{eq:ipfr}
IPFR_m=\frac{1}{i_{m+1}-i_m}\sum_{i=i_m}^{i_{m+1}} \frac{|\mathcal{A}_{i_m}-\mathcal{A}_i|}{\mathcal{A}_{i_m}}
\end{equation}
We compute IPFR only for recovery intervals with at least one intermediate round. Over the full trajectory, this yields $M$ such intervals, and the average IPFR (AIPFR) is:
\begin{equation}
\label{eq:aipfr}
AIPFR=\frac{1}{M}\sum_{m=1}^{M} IPFR_m
\end{equation}
Finally, we define \emph{consistency} as
\begin{equation}
\label{eq:consistency}
C = 1 - AIPFR
\end{equation}
where consistency values close to 1 represent FL algorithms that do not experience drastic drops in accuracy.

\textit{Forgetting on Prior Sampled Clients: } To assess retention of previously encountered clients under partial participation, we adapt backward transfer (BwT) from continual learning \cite{lopez2017gradient}.
Let $K$ be the total number of clients and let $\mathcal{K}=\bigcup_{i=1}^{T-1} K^{i}$ denote the set of clients sampled before round $T$.
For each $k\in\mathcal{K}$, let $t<T$ be the most recent round in which $k$ was sampled.
We define forgetting as
\begin{equation}
\label{eq:forgetting_bwt}
\mathcal{F}=\frac{1}{|\mathcal{K}|}\sum_{k\in\mathcal{K}}\Big(\mathcal{A}_{g,k}^{T}-\mathcal{A}_{g,k}^{t}\Big),
\end{equation}
where $\mathcal{A}_{g,k}^{T}$ is the global model’s accuracy at round $T$ evaluated on client $k$'s test set.
This metric understands how global model updates affect prior sampled clients, as the global model evolves drastically given the current set of sampled clients and does not necessarily retain prior client information.
Therefore, negative values indicate deterioration (forgetting) on previously sampled clients, while positive values indicate net improvement. Overall, these metrics allow for a more rigorous, holistic evaluation of FL algorithms not currently provided in literature.

\input{Tables/table1}

%% file: Tables/table1.tex
\begin{table*}[]
\centering
\caption{Global and local results from partitioning via a Dirichlet distribution $\alpha=0.1$ on BloodMNIST, OrganSMNIST, and OrganCMNIST. Results are averaged and reported across 3 seeds. The best results are bolded.}
\label{tab:tab1}
\resizebox{0.84\textwidth}{!}{%
\begin{tabular}{c|c|cccccc|
>{\columncolor[HTML]{C2F3B2}}c }
\textbf{Dataset}               & \textbf{Strategy}                      & \textbf{Avg Global Acc $\uparrow$} & \textbf{Avg Global BwT $\downarrow$} & \textbf{Avg Global Consistency $\uparrow$} & \textbf{Avg Local Acc $\uparrow$} & \textbf{Avg Worst Client Acc $\uparrow$} & \textbf{Avg Local Consistency $\uparrow$} & \textbf{Balance $\uparrow$} \\ \hline
                               & FedAvg \cite{mcmahan2017communication} & 0.409 ± 0.147                      & -0.255 ± 0.113          & 0.721 ± 0.014                   & 0.775 ± 0.148                     & 0.664 ± 0.024                            & 0.747 ± 0.008                  & 0.592                       \\
                               & FedCurv \cite{shoham2019overcoming}    & 0.472 ± 0.169                      & -0.243 ± 0.122          & 0.733 ± 0.009                   & 0.806 ± 0.141                     & 0.705 ± 0.008                            & 0.742 ± 0.007                  & 0.639                       \\
                               & MOON \cite{li2021model}                & 0.426 ± 0.153                      & -0.251 ± 0.115          & 0.710 ± 0.018                   & 0.787 ± 0.143                     & 0.677 ± 0.009                            & 0.756 ± 0.012                  & 0.606                       \\
                               & FedProx \cite{li2020federated}         & 0.431 ± 0.155                      & -0.280 ± 0.105          & 0.717 ± 0.016                   & 0.828 ± 0.147                     & 0.729 ± 0.004                            & 0.718 ± 0.019                  & 0.629                       \\
                               & FedNTD \cite{lee2022preservation}      & 0.589 ± 0.177                      & -0.191 ± 0.117          & 0.807 ± 0.023                   & 0.849 ± 0.120                     & 0.766 ± 0.008                            & 0.767 ± 0.008                  & 0.719                       \\ \cline{2-9} 
                               & FedAS \cite{yang2024fedas}             & 0.315 ± 0.074                      & -0.260 ± 0.085          & 0.594 ± 0.006                   & 0.773 ± 0.128                     & 0.638 ± 0.038                            & 0.688 ± 0.012                  & 0.544                       \\
                               & FedALA \cite{zhang2023fedala}          & 0.516 ± 0.113                      & -0.225 ± 0.076          & 0.695 ± 0.011                   & \textbf{0.899 ± 0.068}            & \textbf{0.831 ± 0.022}                   & \textbf{0.857 ± 0.024}         & 0.708                       \\
                               & Ditto \cite{li2021ditto}               & 0.450 ± 0.127                      & -0.233 ± 0.069          & 0.696 ± 0.040                   & 0.794 ± 0.120                     & 0.674 ± 0.039                            & 0.710 ± 0.024                  & 0.622                       \\
                               & FedPer \cite{arivazhagan2019federated} & 0.325 ± 0.084                      & -0.286 ± 0.069          & 0.597 ± 0.010                   & 0.782 ± 0.125                     & 0.653 ± 0.048                            & 0.692 ± 0.026                  & 0.554                       \\
                               & FedKD \cite{wu2022communication}       & 0.426 ± 0.102                      & -0.216 ± 0.071          & 0.693 ± 0.044                   & 0.811 ± 0.103                     & 0.686 ± 0.021                            & 0.773 ± 0.017                  & 0.618                       \\
                               & FedBABU \cite{oh2021fedbabu}           & 0.523 ± 0.156                      & -0.248 ± 0.110          & 0.743 ± 0.036                   & 0.870 ± 0.122                     & 0.792 ± 0.002                            & 0.793 ± 0.033                  & 0.696                       \\ \cline{2-9} 
\multirow{-12}{*}{BloodMNIST}  & \textbf{FedKPer (Ours)}                & \textbf{0.645 ± 0.188}             & \textbf{-0.127 ± 0.101} & \textbf{0.823 ± 0.018}          & 0.868 ± 0.122                     & 0.796 ± 0.001                            & 0.796 ± 0.011                  & \textbf{0.757}              \\ \hline
                               & FedAvg \cite{mcmahan2017communication} & 0.445 ± 0.145                      & -0.103 ± 0.069          & 0.811 ± 0.005                   & 0.686 ± 0.137                     & 0.430 ± 0.010                            & 0.809 ± 0.005                  & 0.565                       \\
                               & FedCurv \cite{shoham2019overcoming}    & 0.428 ± 0.143                      & -0.111 ± 0.073          & 0.777 ± 0.009                   & 0.681 ± 0.127                     & 0.429 ± 0.012                            & 0.824 ± 0.000                  & 0.555                       \\
                               & MOON \cite{li2021model}                & 0.445 ± 0.149                      & -0.102 ± 0.068          & 0.783 ± 0.004                   & 0.684 ± 0.140                     & 0.430 ± 0.004                            & 0.796 ± 0.009                  & 0.564                       \\
                               & FedProx \cite{li2020federated}         & 0.453 ± 0.114                      & -0.099 ± 0.051          & 0.826 ± 0.022                   & 0.696 ± 0.106                     & 0.452 ± 0.004                            & 0.837 ± 0.007                  & 0.575                       \\
                               & FedNTD \cite{lee2022preservation}      & 0.514 ± 0.139                      & -0.072 ± 0.059          & 0.851 ± 0.001                   & 0.729 ± 0.135                     & 0.490 ± 0.002                            & 0.798 ± 0.000                  & 0.622                       \\ \cline{2-9} 
                               & FedAS \cite{yang2024fedas}             & 0.294 ± 0.071                      & -0.093 ± 0.055          & 0.651 ± 0.016                   & 0.654 ± 0.103                     & 0.391 ± 0.009                            & 0.745 ± 0.020                  & 0.474                       \\
                               & FedALA \cite{zhang2023fedala}          & 0.375 ± 0.100                      & -0.172 ± 0.056          & 0.719 ± 0.052                   & \textbf{0.750 ± 0.090}            & \textbf{0.584 ± 0.035}                   & 0.757 ± 0.045                  & 0.563                       \\
                               & Ditto \cite{li2021ditto}               & 0.438 ± 0.110                      & -0.089 ± 0.041          & 0.790 ± 0.014                   & 0.675 ± 0.107                     & 0.398 ± 0.011                            & 0.757 ± 0.003                  & 0.557                       \\
                               & FedPer \cite{arivazhagan2019federated} & 0.332 ± 0.077                      & -0.117 ± 0.045          & 0.668 ± 0.043                   & 0.668 ± 0.097                     & 0.424 ± 0.014                            & 0.767 ± 0.028                  & 0.500                       \\
                               & FedKD \cite{wu2022communication}       & 0.224 ± 0.073                      & -0.182 ± 0.039          & 0.650 ± 0.018                   & 0.569 ± 0.088                     & 0.372 ± 0.042                            & 0.662 ± 0.093                  & 0.396                       \\
                               & FedBABU \cite{oh2021fedbabu}           & 0.459 ± 0.118                      & -0.095 ± 0.059          & 0.806 ± 0.013                   & 0.739 ± 0.111                     & 0.491 ± 0.005                            & 0.823 ± 0.008                  & 0.599                       \\ \cline{2-9} 
\multirow{-12}{*}{OrganSMNIST} & \textbf{FedKPer (Ours)}                & \textbf{0.517 ± 0.130}             & \textbf{-0.024 ± 0.037} & \textbf{0.899 ± 0.015}          & 0.736 ± 0.111                     & 0.511 ± 0.003                            & \textbf{0.862 ± 0.011}         & \textbf{0.626}              \\ \hline
                               & FedAvg \cite{mcmahan2017communication} & 0.568 ± 0.207                      & -0.077 ± 0.117          & 0.818 ± 0.017                   & 0.844 ± 0.131                     & 0.717 ± 0.001                            & 0.747 ± 0.009                  & 0.706                       \\
                               & FedCurv \cite{shoham2019overcoming}    & 0.551 ± 0.201                      & -0.081 ± 0.125          & 0.806 ± 0.009                   & 0.830 ± 0.126                     & 0.691 ± 0.006                            & 0.803 ± 0.002                  & 0.690                       \\
                               & MOON \cite{li2021model}                & 0.559 ± 0.203                      & -0.082 ± 0.115          & 0.805 ± 0.003                   & 0.840 ± 0.131                     & 0.712 ± 0.005                            & 0.792 ± 0.003                  & 0.700                       \\
                               & FedProx \cite{li2020federated}         & 0.544 ± 0.146                      & -0.073 ± 0.090          & 0.847 ± 0.015                   & 0.836 ± 0.107                     & 0.701 ± 0.001                            & 0.768 ± 0.015                  & 0.690                       \\
                               & FedNTD \cite{lee2022preservation}      & 0.618 ± 0.175                      & -0.066 ± 0.094          & 0.838 ± 0.007                   & 0.868 ± 0.109                     & 0.754 ± 0.004                            & 0.859 ± 0.010                  & 0.743                       \\ \cline{2-9} 
                               & FedAS \cite{yang2024fedas}             & 0.371 ± 0.085                      & -0.162 ± 0.080          & 0.685 ± 0.002                   & 0.810 ± 0.087                     & 0.656 ± 0.009                            & 0.794 ± 0.010                  & 0.591                       \\
                               & FedALA \cite{zhang2023fedala}          & 0.523 ± 0.107                      & -0.129 ± 0.060          & 0.809 ± 0.040                   & 0.884 ± 0.065                     & 0.803 ± 0.026                            & 0.807 ± 0.063                  & 0.703                       \\
                               & Ditto \cite{li2021ditto}               & 0.564 ± 0.131                      & -0.113 ± 0.060          & 0.811 ± 0.034                   & 0.841 ± 0.094                     & 0.716 ± 0.015                            & 0.819 ± 0.024                  & 0.702                       \\
                               & FedPer \cite{arivazhagan2019federated} & 0.415 ± 0.096                      & -0.174 ± 0.071          & 0.711 ± 0.033                   & 0.818 ± 0.082                     & 0.682 ± 0.011                            & 0.813 ± 0.005                  & 0.616                       \\
                               & FedKD \cite{wu2022communication}       & 0.328 ± 0.083                      & -0.179 ± 0.067          & 0.692 ± 0.006                   & 0.731 ± 0.091                     & 0.570 ± 0.070                            & 0.741 ± 0.002                  & 0.530                       \\
                               & FedBABU \cite{oh2021fedbabu}           & 0.557 ± 0.146                      & -0.090 ± 0.113          & 0.836 ± 0.011                   & 0.855 ± 0.093                     & 0.732 ± 0.006                            & 0.835 ± 0.008                  & 0.706                       \\ \cline{2-9} 
\multirow{-12}{*}{OrganCMNIST} & \textbf{FedKPer (Ours)}                & \textbf{0.660 ± 0.134}             & \textbf{-0.013 ± 0.035} & \textbf{0.924 ± 0.006}          & \textbf{0.889 ± 0.092}            & \textbf{0.797 ± 0.004}                   & \textbf{0.868 ± 0.008}         & \textbf{0.774}         \\ \hline    
\end{tabular}%
}
\end{table*}

%% file: Sections/4_results.tex
\section{Results}
\subsection{Experimental Design}
We evaluate FedKPer on three medical datasets: BloodMNIST, OrganCMNIST, and OrganSMNIST \cite{yang2023medmnist}. For BloodMNIST, OrganCMNIST, and OrganSMNIST we create a total of $20, 30,$ and $50$ local clients respectively. Local clients are created by partitioning via a Dirichlet distribution ($\alpha=0.1$) as done in \cite{li2021model} for a high level of heterogeneity. At each round, $10\%$ of clients are sampled. We utilize the standard four-layer CNN common in FL literature and a SGD optimizer with $\eta =0.01$ \cite{mcmahan2017communication, lee2022preservation}. 
Each local model is trained for $5$ epochs, with the total communication rounds set to $100$.

For each dataset, we reserve the overall test set, denoting this as the global test set to measure generalizability of the global model. Additionally, each local client $k$ has its own distinct dataset $\mathcal{D}_{k}$. From $\mathcal{D}_{k}$, client $k$'s local training and test set is derived, where $20\%$ of $\mathcal{D}_{k}$ is reserved for local testing. Besides the metrics reported in Section~\ref{subsec:metrics}, we report average worst-client local accuracy, along with mean global and mean local accuracy. We also devise a measure of global-local balance to understand how an algorithm handles both tasks, obtained via $\frac{\mathcal{\bar{A}}_g + \mathcal{\bar{A}}_k}{2}$ where $\mathcal{\bar{A}}_g$ and $\mathcal{\bar{A}}_k$ represent the average global and local test set accuracies respectively.

\subsection{Experimental Results}
In Table \ref{tab:tab1}, we compare FedKPer against a range of classical FL and pFL baselines. FedKPer achieves strong generalization and personalization, leading global accuracy on all three datasets. On BloodMNIST, for instance, it improves average global test accuracy by $9.8\%$ over the next best method. It also better retains prior knowledge, reflected by improved BwT and consistency. FedKPer is also highly competitive with pFL approaches in its personalization capabilities. Notably, despite using local training accuracy as a lightweight reliability proxy in aggregation, FedKPer achieves best or second-best worst-client accuracy across datasets, indicating reduced client performance gaps and avoiding undue influence from a subset of clients via the combined reliability-diversity weighting.
Overall, FedKPer delivers the strongest \textbf{balance} (final column of Table~\ref{tab:tab1}) without sacrificing retention.
\begin{figure}[h]
    \centering
    \begin{subfigure}[b]{0.49\linewidth}
        \centering
        \includegraphics[width=\linewidth]{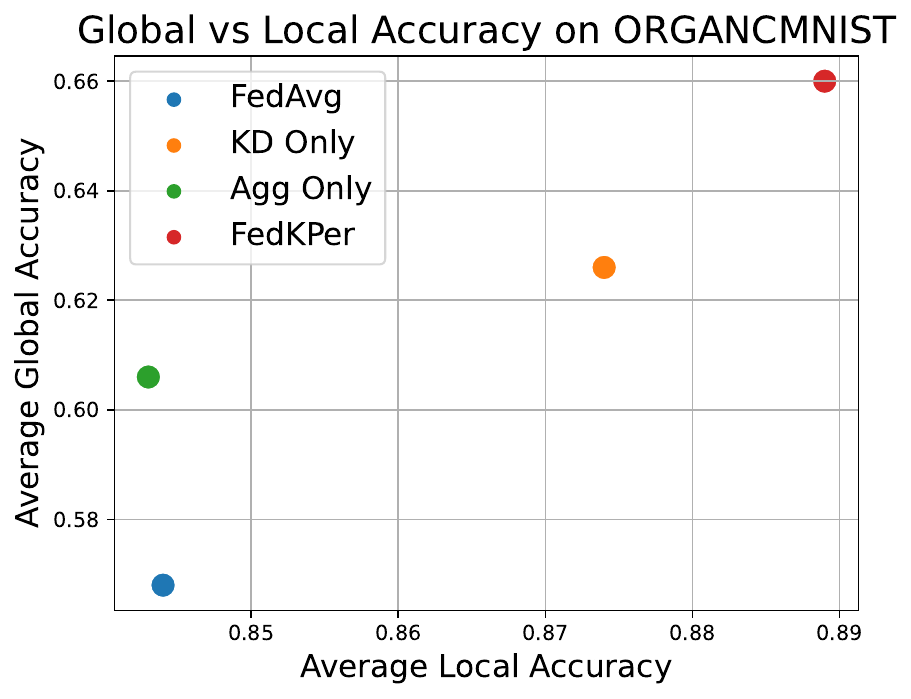}
        \caption{}
    \end{subfigure}
    \begin{subfigure}[b]{0.50\linewidth}
        \centering
        \includegraphics[width=\linewidth]{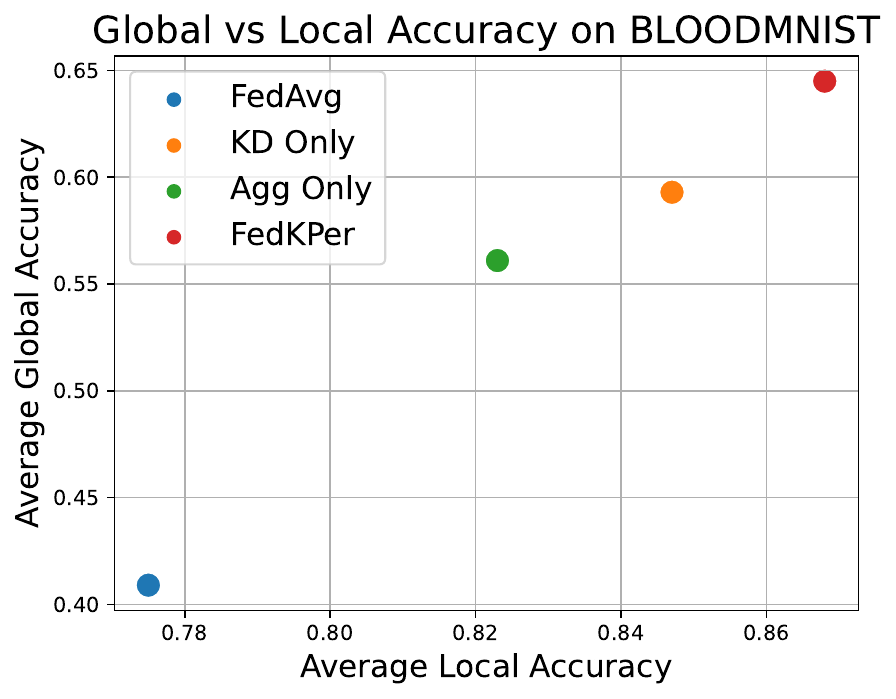}
        \caption{}
    \end{subfigure}
    \caption{FedKPer component comparison: KD Only is Equation~\ref{eqn_loss}, whereas Agg Only is Equation~\ref{eqn_score}.}
    \label{fig:effect_of_comp}
\end{figure}
\begin{figure}[h]
    \centering
    \includegraphics[width=0.34\textwidth]{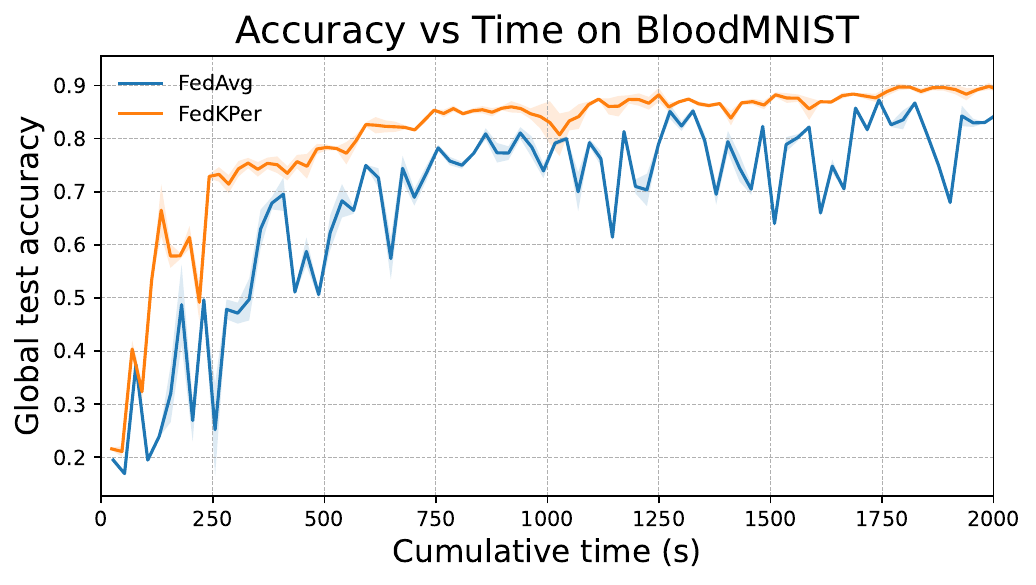}
    \caption{Budgeted efficiency between FedAvg and FedKPer}
    \label{fig:acc_vs_time}
\end{figure}

Additionally, we wish to understand and isolate the effect of each component of FedKPer. In Figure~\ref{fig:effect_of_comp}, we compare global and local performance of FedKPer, FedAvg, the aggregation component of FedKPer (Equation~\ref{eqn_score}), and the adaptive distillation loss (Equation~\ref{eqn_loss}). As indicated by the figure, the two components are complementary, as the adaptive distillation term primarily boosts local performance, while the proposed aggregation primarily improves global performance, and combining both yields the best overall trade-off. 

Finally, to assess practical compute efficiency, we plot global accuracy as a function of cumulative wall-clock time (3-seed mean) against FedAvg under identical hardware, training, and evaluation settings. FedKPer achieves higher accuracy at matched time budgets; for BloodMNIST, at 500s it improves global accuracy by $38.8\%$. Overall, FedKPer improves the global-local trade-off while delivering better accuracy under realistic training budgets.


%% file: Sections/5_conclusion.tex
\section{Conclusion}
We propose FedKPer, a federated learning method that better balances personalization and generalization under statistical heterogeneity while reducing forgetting. FedKPer adaptively controls each client’s alignment with the global model to preserve useful shared knowledge while enabling local adaptation, and uses a reliability- and label-diversity-aware aggregation to improve generalization. We also introduce forgetting-oriented evaluation metrics for a more complete assessment of FL methods. Across three medical imaging datasets, FedKPer improves the personalization-generalization trade-off with lower forgetting.
